\documentclass[aps,pra,floatfix,superscriptaddress,twocolumn,showpacs,10pt]{revtex4-1} \usepackage{amssymb, amsmath,color,mciteplus,graphicx,subfigure,amsthm}
\usepackage{times}

\begin{document}

\title{Further Investigation on Classical Multiparty computation using Quantum Resources}

\author{Hao Cao } \email{caohao2000854@163.com}
\affiliation{State Key Laboratory of Integrated Service Networks, Xidian University, Xi'an, China 710071}
\affiliation{School of Information and Network Engineering, Anhui Science and Technology University, Chuzhou, China 233100}

\author{ Wenping Ma} \email{wp_ma@mail.xidian.edu.cn}
\affiliation{State Key Laboratory of Integrated Service Networks, Xidian University, Xi'an, China 710071}

\author{Ge Liu} \email{446552402@qq.com}
\affiliation{State Key Laboratory of Integrated Service Networks, Xidian University, Xi'an, China 710071}

\author{Liangdong Lyu} \email{kelinglv@163.com}
\affiliation{State Key Laboratory of Integrated Service Networks, Xidian University, Xi'an, China 710071}
\affiliation{Department of Basic Sciences, Air Force Engineering University, Xi'an, 710071, China}

\date{\today}

\begin{abstract}
The tremendous development of cloud computing and network technology makes it possible for multiple people with limited resources to complete a large-scale computing with the help of cloud servers. In order to protect the privacy of clients, secure multiparty computation plays an important role in the process of computing. Recently, Clementi et al[\textcolor[rgb]{0.00,0.07,1.00}{Phys. Rev. A {\bf 96}, 062317(2017)}] proposed a secure multiparty computation protocol using quantum resources. In their protocol, utilizing only linear classical computing and limited manipulation of quantum information, a method of computing $n-variable$ symmetric Boolean function $f(x_1, x_2, \cdots, x_n)$ with degree 2 is proposed, and all clients can jointly compute $f(x_1, x_2, \cdots, x_n)$  without revealing their private inputs with the help of a sever. They proposed an open problem: are there more simple nonlinear functions like the one presented by them that can be used as subroutines for larger computation protocols?  We will give the answer to this question in this paper. Inspired by Clementi et al's work, we continue to explore the quantum realization of Boolean functions. First, we demonstrate a way to compute a class of $n-variable$ symmetric Boolean function $f_n^k$ by using single-particle quantum state $|0\rangle$ and single-particle unitary operations $U_k$. Second, we show that each $n-variable$ symmetric Boolean function can be represented by the linear combination of $f_n^k(k=0,1,\cdots,n)$ and each function $f_n^k(2\leq k\leq n)$ can be used to perform secure multiparty computation. Third, we propose an universal quantum implementation method for arbitrary $n-variable$ symmetric Boolean function $f(x_1, x_2, \cdots, x_n)$. Finally, we demonstrate our secure multiparty computation protocol on IBM quantum cloud platform.
\end{abstract}


\maketitle

\section{Introduction}
In the cloud environment, it is very common for a number of clients with limited resource to delegate the server to compute a function of their inputs. If each client would rather not reveal his input information to server and other clients, a special cryptographic model called Secure multiparty computation(SMPC) will be considered for use. The SMPC problem is originated from the Yao's millionaire problem\cite{Y1982}, and many classical solutions\cite{Y1982,Y1986,OS1987} to it have been proposed.

In 1984, a quantum key distribution protocol known as BB84\cite{BB1984}, which is completely different from classical cryptography, came into people's vision and attracted wide attention. Since then, various types of quantum cryptographic protocols, such as quantum secret sharing(QSS)\cite{HBB1999,CM2017}, quantum secure direct communication(QSDC)\cite{ZDS2017}, quantum key agreement(QKA)\cite{CM2018}, quantum privacy comparison(QPC)\cite{TLH2012}, and so on, have been proposed. Especially, quantum solutions to SMPC problem, i.e., quantum SMPC(QSMPC) \cite{AB2009,LB2010,DKK2016,BDS2016,CPE2017}, attracts much attention because of its widely application in electronic voting, online auction, and multiparty data processing.

Recently, Clementi et al\cite{CPE2017} proposed a QSMPC protocol in which a number of clients can collaborate to compute a function without revealing their inputs. The function in their protocol is a $n-variable$ symmetric Boolean function with degree 2. They proposed an open problem: are there more simple nonlinear functions like the one presented by them that can be used as subroutines for larger computation protocols? We will give the answer to this question in this paper. Inspired by Clementi et al's work, we focus on the case of arbitrary symmetric Boolean functions. First, we give a quantum implementation of a class of symmetric Boolean function $f_n^k$. Second, we show that each $n-variable$ symmetric Boolean function can be represented by linear combination of $f_n^k(k=0,1,\cdots,n)$ and each function $f_n^k$ can be used to perform secure multiparty computation. Third, we explore an universal quantum implementation method for arbitrary $n-variable$ symmetric Boolean function $f(x_1, x_2, \cdots, x_n)$.

The remainder of our work is organized as follows. Section II introduces a method to compute a a class of $n-variable$ symmetric Boolean function $f_n^k$ by using single-particle quantum state $|0\rangle$ and single-particle unitary operations $U_k$. Besides, the quantum implementations of arbitrary symmetric Boolean functions are explored in this section. In section III, we give the description of our QSMPC protocol, i.e., each function $f_n^k(2\leq k\leq n)$ can be used to perform secure multiparty computation. Section IV analyzes the security of our protocol. In section V, a simulation of our protocol on IBM quantum cloud platform. At last, a conclusion of this paper is given in section VI.

\section{Theory}
Our work is a further investigation of Clementi et al's work\cite{CPE2017}. In their paper, they focus on computing the $n-variable$ Boolean function $f$ by using the following equations:
\begin{equation}\label{quantum implementation 1 of $f^n_2$}
\begin{array}{l}
  (U^\dag)^{\oplus_i x_i}U^{x_n}\cdots U^{x_2}U^{x_1}|0\rangle=|f\rangle,
 \end{array}
 \end{equation}
and
\begin{equation}\label{quantum implementation 2 of $f^n_2$}
\begin{array}{l}
  (U^\dag)^{\oplus_i x_i}V^{r_n}U^{x_n}\cdots V^{r_2}U^{x_2}V^{r_1}U^{x_1}|0\rangle=|\bar{r}\oplus f\rangle,
 \end{array}
 \end{equation}
where $U=R_y(\frac{\pi}{2})=cos\frac{\pi}{4}I-isin\frac{\pi}{4}Y$ be the $\frac{\pi}{2}$ rotation around the $y$ axis of the Bloch sphere, $V=R_y(\pi)=cos\frac{\pi}{2}I-isin\frac{\pi}{2}Y$ be the $\pi$ rotation around the $y$ axis, $f$ be the $2-degree$ $n-variable$ symmetric Boolean function $f_2^n=\oplus_{j=1}^{n-1}[x_{j+1}\times(\oplus_{i=1}^{j}x_i)]=\oplus_{1\leq i < j \leq n}x_ix_j$, $r_i\in \{0, 1\}(i=1,2,\cdots,n)$ and $\bar{r}=\oplus_{i=1}^{n}r_{i}$.

Define $U_k=R_y(\frac{\pi}{k}) $ $(k = 1, 2, \cdots, n)$ be the $\frac{\pi}{k}$ rotation around the $y$ axis of the Bloch sphere, and $U_0=I=R_y(0) $ be the identity operation.  Then we have  the following important arguments about $U_k$ $(k = 1, 2, \cdots, n)$.

 {\bf Theorem 1}~~ Let $k$ and $h$ be two nonnegative integers, then the following equation holds:
  \begin{equation}\label{Proof of Theorem 2 in $U_k$ and $U_h$}
\begin{array}{ll}
 U_k U_h= U_h U_k ;  &  U_k^\dag U_h= U_h U_k^\dag.
 \end{array}
 \end{equation}
  \begin{proof}
  ~~First, $U_k U_h$ $ = $ $R_y(\frac{\pi}{k})R_y(\frac{\pi}{h})$ $ = $ $R_y(\frac{\pi}{k}+\frac{\pi}{h})$ $ = $ $R_y(\frac{\pi}{h}+\frac{\pi}{k})$ $ = $ $R_y(\frac{\pi}{h})R_y(\frac{\pi}{k})$ $ = $ $ U_h U_k$.

  Next,  $U_h$ $ = $ $I U_h$ $ = $ $(U_k^\dag U_k) U_h$ $\Rightarrow $ $U_h$ $ = $ $U_k^\dag U_h U_k$ $\Rightarrow $ $U_hU_k^\dag$ $ = $ $U_k^\dag U_h$.
 \end{proof}

 {\bf Theorem 2} ~~Let $k$ be a nonnegative integer, $x_i \in \{0, 1\}$ $(i=1, 2, \cdots, n)$.

(1)~The single-particle quantum state in the form of $(U_k^\dag)^{(\sum_ix_i) \mod k}U_k^{x_n}\cdots U_k^{x_2}U_k^{x_1}|0\rangle$ can be regarded as the quantum implementation of a $n-variable$ Boolean function $f_n^k$, i.e.,
\begin{equation}\label{quantum implementation 1 of $f^n_k$ 2}
\begin{array}{l}
  (U_k^\dag)^{(\sum_ix_i) \mod k}U_k^{x_n}\cdots U_k^{x_2}U_k^{x_1}|0\rangle = |f_n^k \rangle.
 \end{array}
 \end{equation}

 (2)~$f_n^k$ is a symmetric Boolean function.

 (3)~$deg(f_n^k)\geq k$.

 (4)~The algebraic normal form of $f_n^k$ contains all monomials with degree $k$.

 (5)~Each  $n-variable$ symmetric Boolean function can be represented by the linear combination of $f_n^k(k=0,1,\cdots,n)$.
 \begin{proof}
~~(1)~In order to prove the proposition (1), we only need to show that the quantum state in the form of $(U_k^\dag)^{(\sum_ix_i) \mod k}U_k^{x_n}\cdots U_k^{x_2}U_k^{x_1}|0\rangle$ is either $|0\rangle$ or $|1\rangle$. Let $wt(x_1, x_2, \cdots, x_n)=\sum_ix_i=ak+b$, where $wt(x_1, x_2, \cdots, x_n)$ be the weight of $(x_1, x_2, \cdots, x_n)$(i.e., the number of ones in $x_1, x_2, \cdots, x_n$), $a$ and $b$ be two nonnegative integer, and $0 \leq b < k$, then
\begin{equation}\label{Proof of Theorem 1 in $f^n_k$ 2}
\begin{array}{ll}
  &(U_k^\dag)^{(\sum_ix_i) \mod k}U_k^{x_n}\cdots U_k^{x_2}U_k^{x_1}|0\rangle\\[2mm]
 =&(U_k^\dag)^{(\sum_ix_i) \mod k}U_k^{\sum_ix_i}|0\rangle \\[2mm]
 =&(U_k^\dag)^{b}U_k^{ak+b}|0\rangle \\[2mm]
 =&(U_k)^{ak}|0\rangle \\[2mm]
 =&[R_y(\frac{\pi}{k})]^{ak}|0\rangle \\[2mm]
 =&R_y(a\pi)|0\rangle \in \{|0\rangle, |1\rangle\},\\[2mm]
 \end{array}
 \end{equation}
  Hence, $(U_k^\dag)^{(\sum_ix_i) \mod k}U_k^{x_n}\cdots U_k^{x_2}U_k^{x_1}|0\rangle$ can be regarded as the quantum implementation of a $n-variable$ Boolean function $f_n^k$.

 (2)~ From equation(\ref{Proof of Theorem 1 in $f^n_k$ 2}), we can easily get that $f_n^k=1$ if and only if $R_y(a\pi)|0\rangle =|1\rangle$, if and only if  $a$ be odd. Let $a^\prime$ be a nonnegative integer, we have
 $$
 f_n^k= \begin{cases}  0 & wt(x_1, x_2, \cdots, x_n)=2a^\prime k+b\\[2mm]
                       1 & wt(x_1, x_2, \cdots, x_n)=(2a^\prime+1)k+b \end{cases}.
 $$
 which implies that $f_n^k$ is a symmetric Boolean function.

(3)~It is easy to very the fact that $f_n^k(x_1, x_2, \cdots, x_n)=0$ for each vector $(x_1, x_2, \cdots, x_n)$ with $wt(x_1, x_2, \cdots, x_n) < k$, which implies $deg(f_n^k)\geq k$.

(4)~Owing to the fact that $f_n^k(x_1, x_2, \cdots, x_n)=1$ for each vector $(x_1, x_2, \cdots, x_n)$ with $wt(x_1, x_2, \cdots, x_n) = k$, we get that the algebraic normal form of $f_n^k$ contains all monomials with degree $k$.

(5)~From (3) and (4), we can draw that $f_n^0, $ $f_n^1, $ $f_n^2, $ $\cdots, $ and $f_n^n $ are linearly independent, and they from a basis of $n-variable$ symmetric Boolean functions. Hence, each  $n-variable$ symmetric Boolean function can be represented by the linear combination of $f_n^k(k=0,1,\cdots,n)$.
 \end{proof}

 From {\bf Theorem 1} and {\bf Theorem 2}, we can easily get the following equation which is the generalization of equation(\ref{quantum implementation 2 of $f^n_2$}).
 \begin{equation}\label{quantum implementation 10 of $f^n_k$}
\begin{array}{l}
  (U^\dag)^{(\sum_ix_i) \mod k}V^{r_n}U^{x_n}\cdots V^{r_1}U^{x_1}|0\rangle=|\bar{r}\oplus f_n^k\rangle,
 \end{array}
 \end{equation}
 where $r_i\in \{0, 1\}(i=1,2,\cdots,n)$ and $\bar{r}=\oplus_{i=1}^{n}r_{i}$.

 For convenience of description, we denote the vector $(x_1, x_2, \cdots, x_1)$ as $x$, the vector $(r_1, r_2, \cdots, r_1)$ as $r$, the unitary operation $(U_k^\dag)^{(\sum_ix_i) \mod k}$ $U_k^{x_n}$ $\cdots$ $U_k^{x_2}$ $ U_k^{x_1}$ as ${U}(n,k,x)$, and the unitary operation $(U^\dag)^{(\sum_ix_i) \mod k}$ $V^{r_n}U_k^{x_n}$ $\cdots$ $ V^{r_2}$ $U_k^{x_2}$ $V^{r_1}$ $U_k^{x_1}$ as ${VU}(n,k,x,r)$ separately, then equation(\ref{quantum implementation 1 of $f^n_k$ 2}) and equation(\ref{quantum implementation 10 of $f^n_k$}) can be rewritten as follows:

 \begin{equation}\label{quantum implementation 3 of $f^n_k$}
\begin{array}{l}
  {U}(n,k,x)|0\rangle=|f_n^k\rangle\\[2mm]
 \end{array}
 \end{equation}
  \begin{equation}\label{quantum implementation 7 of $f^n_k$}
\begin{array}{l}
  {VU}(n,k,x,r)|0\rangle=| \bar{r} \oplus f_n^k\rangle
 \end{array}
 \end{equation}

{\bf Theorem 3} ~~Let $k$ and $h$ be two nonnegative integers, then
 \begin{equation}\label{quantum implementation 1 of $f^n_k$ and $f^n_h$}
 \begin{array}{cl}
   |f^n_k\oplus f^n_h\rangle =&\widetilde{U}(n,k,x)\widetilde{U}(n,h,x)|0\rangle
 \end{array}
 \end{equation}

   \begin{proof}
   First, let $r_n=r_{n-1}=\cdots r_2=0$ and $r_1=1$ in
    equation (\ref{quantum implementation 10 of $f^n_k$}), we get
 \begin{equation}\label{quantum implementation 4 of $f^n_k$}
 \begin{array}{cl}
 (U_k^\dag)^{(\sum_ix_i) \mod k} U_k^{x_n} \cdots U_k^{x_2}U_k^{x_1} |1\rangle=|f_n^k \oplus 1\rangle
  \end{array}
 \end{equation}
 i.e.,
  \begin{equation}\label{quantum implementation 5 of $f^n_k$}
 \begin{array}{cl}
 \widetilde{U}(n,k,x) |1\rangle=|f_n^k \oplus 1\rangle
  \end{array}
 \end{equation}
  ~~Next, consider the right hand of the equation(\ref{quantum implementation 1 of $f^n_k$ and $f^n_h$}).
 \begin{equation}\label{quantum implementation 6 of $f^n_k$}
 \begin{array}{cl}
   &\widetilde{U}(n,k,x)\widetilde{U}(n,h,x)|0\rangle \\[2mm]
  =&\widetilde{U}(n,k,x)|f_n^h\rangle \\[2mm]
  =&\begin{cases}  |f_n^k\rangle & if~~f_n^h=0\\[2mm] |f_n^k \oplus 1\rangle & if~~f_n^h=1 \end{cases}~~~(From~~ equation(\ref{quantum implementation 4 of $f^n_k$}))\\[2mm]
    =&|f_n^k\oplus f_n^h\rangle\\[2mm]
 \end{array}
 \end{equation}
 \end{proof}
 From {\bf Theorem 2(5)} and {\bf Theorem 3}, we can easily get the quantum implementation of an arbitrary symmetric Boolean function.

{\bf Theorem 4} ~~Let $f(x_1, x_2, \cdots, x_n)$ be a $n-variable$ symmetric Boolean function which can be presented as
 \begin{equation}\label{representation 3 of symmetric boolean function}
 \begin{array}{cl}
  f=\oplus_{k=0}^na_kf_n^k
 \end{array}
 \end{equation}
 then
 \begin{equation}\label{representation 4 of symmetric boolean function}
 \begin{array}{cl}
  |f\rangle=\prod _{k=0}^n [\widetilde{U}(n,k,x)]^{a_k}|0\rangle
 \end{array}
 \end{equation}

\section{Quantum Secure Multiparty Computation Protocol}
In this section, we will show the QSMPC protocol by using
equation (\ref{quantum implementation 10 of $f^n_k$})
or equation (\ref{quantum implementation 7 of $f^n_k$}) in the ideal case  environment. Let $C_i$ $(i=1, 2, \cdots, n)$ be $n$ clients and each client $C_i$ possesses a private input $x_i$ and selects a random bit $r_i$. They want to jointly compute the function $f^n_k(2\leq k\leq n)$ without revealing their inputs with the help of a server $S$.

{\bf Step 1~} First, each client $C_i$ divides his private input $x_i$ into $n$ elements $x_{i,1}, $ $x_{i,2}, $ $\cdots, $ and $x_{i,n} $ $(x_{i,j}\in \{0,1,\cdots,k-1\}, j=1,2,\cdots,n)$ , and the random bit $r_i$ into $n$ elements $r_{i,1}, $ $r_{i,2}, $ $\cdots, $ and $r_{i,n} $ $(r_{i,j}\in \{0,1\}, j=1,2,\cdots,n)$ ,  such that $\sum_{j=1}^n x_{i,j}\equiv x_i \mod k $ and $\oplus_{j=1}^n r_{i,j}= r_i  $. Second, each client $C_i$ sends $x_{i,j} $ and $r_{i,j} $ to client $C_j$ . Third, each client $C_i$ computes $\tilde{x}_i=\sum_{j=1}^n x_{j,i} \mod k $ and $\tilde{r}_i=\oplus_{j=1}^n r_{j,i} $.

{\bf Step 2~} First, the server $S$ prepares a single particle in the state $|0 \rangle$ and sends it to the client $C_1$. Second, $C_1$ performs the unitary operation $V^{r_1}U^{x_1}$ on the received single particle according to his private input $x_1$ and random bit $r_i$, and sends the resulted single particle to the client $C_2$ who will perform the unitary operation $V^{r_2}U^{x_2}$ on the received particle according to his private input $x_2$ and random bit $r_2$. This procession continues until all the clients have applied their unitary operations to the single particle. At this point, the single particle is in the hands of $C_n$.

{\bf Step 3~} First, each client $C_i$ sends $\tilde{x}_i$ to the client $C_n$ through an secure authentication channel. Second, $C_n$ calculates $(\sum _{i=1}^n \tilde{x}_i)\mod k $ . Third, $C_n$ perform the unitary operation $(U^\dag)^{(\sum _{i=1}^n \tilde{x}_i)\mod k}$ on the single particle, and the resulted particle will be in the state of $|f_n^k \oplus \bar{r} \rangle$ owing to the fact that $(\sum _{i=1}^n \tilde{x}_i)\mod k $ $=$ $(\sum _{i=1}^n x_i)\mod k $. Fourth, $C_n$ sends the resulted particle back to the server $S$ . At last, the server $S$ will get the state $|f_n^k \oplus \bar{r}\rangle$  by measuring the received particle, and announce $f_n^k \oplus \bar{r}$ to all clients.

{\bf Step 4~} First, each client $C_i$ transmits the classical bit $\tilde{r}_i=\oplus_{j=1}^n r_{j,i} $ to all other clients through an secure authentication channel. Second, each client $C_i$ calculates $\oplus_{j=1}^n \tilde{r}_j=\oplus_{j=1}^n \tilde{r}_j=\bar{r}$. At last, every client will extract the value of $f_n^k$ by performing the XOR operation $f_n^k=(f_n^k \oplus \bar{r})\oplus \bar{r}$.

\section{Security Analysis and Efficiency Comparison of the Proposed  QSMPC Protocol }

\subsection{Security Analysis}
In this section, we only focus on the internal attack from the clients or the server because  internal attacks are usually more effective than external attacks. We assume that both the clients and the server will execute the QSMPC protocol. However, the server $S$ will try to get the private inputs of the clients or the function output $f_n^k$, and each client will also try to get the private inputs of other clients.

(1)~Consider the security against the attack from the server $S$ . First, if the server $S$ wants to extract the private input of some one client, say $C_1$, he must intercept the particle sent from $C_1$ to $C_2$, and measures it in the correct measurement basis $\{|0\rangle, |1\rangle\}$ or $\{U_k|0\rangle, U_k|1\rangle\}$. However, he could not choose the right measurement base because he knows nothing about the unitary operation $V^{r_1}U^{x_1}$ and the state information of the single particle. Second, if the server $S$ wants to extract the function output $f_n^k$. However, he will also fail because he knows nothing about the random classical bits $r_i(i=1, 2, \cdots, n)$, $\tilde{r}_i(i=1, 2, \cdots, n)$ and $\bar{r}$.

(2)~Consider the security against the attack from the clients. Let us discuss a very unfavorable situation, i.e., the dishonest clients consist of $C_n$ who will get the information of $(\sum _{i=1}^n \tilde{x}_i)\mod k $, and some other $n-t-1$ clients $C_{t+1}, \cdots, C_{n-1}$. They will collaborate to extracts the private inputs of $C_1$, $\cdots $, $C_t$. Apparently, they  could easily access the $(\sum _{i=1}^t \tilde{x}_i)\mod k $ from $(\sum _{i=1}^n \tilde{x}_i)\mod k $ and $x_{t+1}, \cdots, x_{n-1}, x_{n}$. If $t=1$ (i.e., only one client is honest), they could extract the private input $x_1$; or else, they cannot get any useful information about the private inputs of the honest clients.

\subsection{Efficiency Comparison}
In classical case, secure multi-party computation is usually achieved by garbled circuits\cite{HEK2011}. Similarly, quantum circuits are usually used to perform quantum secure multi-party computation and a quantum circuit consists of a number of quantum gates. The efficiency evaluation usually includes three indicators: the quantum resource, the size and the number of quantum gates, and classical communication cost involved in the quantum circuit.

Several quantum implementations of classical NAND gate, which are based on either entangled GHZ state or single qubit, are presented in the previous work\cite{DKK2016}. In the procession of performing NAND gate based on single qubit, one qubit, two single-particle unitary operations and several rounds of classical communication are involved. In the procession of performing NAND gate based on single entangled GHZ state, three qubit, Four single-particle unitary operations, and several rounds of classical communication are involved. Besides, their protocols guarantees no security for the inputs of the parties as stated in Clementi et al's work\cite{CPE2017}. Quantum implementation of Boolean function and its application in QSMPC are explored in Clementi et al's work\cite{CPE2017}. In their protocol, one qubit, $n+1$ single-particle unitary operations and one round classical communication are needed if there are $n$ clients are involved . Hence, their protocol is more efficient.

Our QSMPC protocol is a generalization of the work by Clementi et al\cite{CPE2017}. From the processions of Clementi et al's protocol and ours, we can know that the efficiency of the two protocols is the same.

\section{Quantum simulation on IBM quantum cloud platform}
We simulates our protocol on IBM quantum cloud platform. Here, we will design the quantum circuits to fit the ibmqx4 quantum computer. Suppose the preparation and measurement of the qubit are operated by the server(i.e., q[0], q[1],  q[2], q[3] and q[4]). The private inputs of $n$ clients and the random classical bits selected by them can be represented as  $x=(x_1, x_2, \cdots, x_n)$ and  $r=(r_1, r_2, \cdots, r_n)$.  Fig.1 shows the quantum circuits of our protocol  with $n=5$, $n=8$, and $n=10$. The definition of the operation $VU(n,k,x,r)$ can be seen in equation(\ref{quantum implementation 7 of $f^n_k$}). The unitary operation $U_k$ is realized by the quantum gate $U_3(\frac{\pi}{k},0,0)$, the unitary operation $V$ is realized by the quantum gate $U_3(\pi,0,0)$ and the unitary operation $I$ is realized by the quantum gate $id$ instead of $U_3(0,0,0)$ in the IBM quantum cloud platform.

\begin{figure}[tbp]\centering
\includegraphics[height=14cm, width=\textwidth]{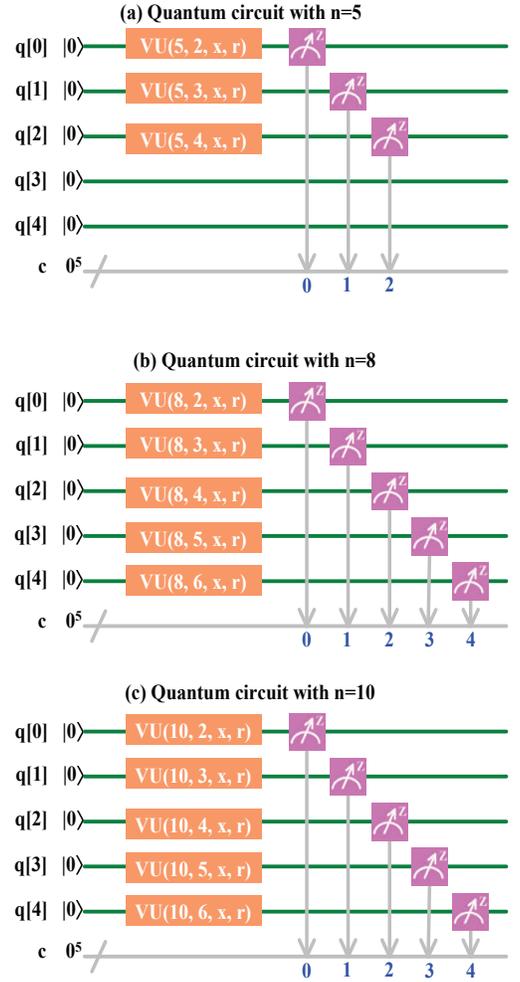}
\caption{(Color online) Quantum implementation of our protocol with $n=5$, $n=8$, and $n=10$. (a) The quantum circuit will output ($f_5^4$ , $f_5^3$, $f_5^2$) by measuring $q[2], $ $q[1]$ and $q[0]$. (b) The quantum circuit will output ($f_8^6$ , $f_8^5$, $f_8^4$, $f_8^3$, $f_8^2$) by measuring $q[4], $ $q[3]$, $q[2], $ $q[1]$ and $q[0]$. (c) The quantum circuit will output ($f_{10}^6$ , $f_{10}^5$, $f_{10}^4$, $f_{10}^3$, $f_{10}^2$) by measuring $q[4], $ $q[3]$, $q[2], $ $q[1]$ and $q[0]$.}
\label{Visio-secure multiparty computation with n=5}
\end{figure}

Case 1: $n=5$. let $C_1, $ $C_2, $ $\cdots, $ and $C_5$ be the five clients. They will collaborate to compute the function $f_5^2,$ $f_5^3,$ and $f_5^4$ with  $x=(x_1, x_2, \cdots, x_5)$ $=$ $(1, 1, 0,  0, 1)$, $r=(r_1, r_2, \cdots, r_5)$ $=$ $(1, 0, 0,  1, 0)$, $(\sum_i^5x_i) \mod 2=1$, $(\sum_i^5x_i) \mod 3=0$,   $(\sum_i^5x_i) \mod 4=3$ and $\bar{r}=\oplus_{i=1}^5r_i=0$.  The quantum implementation of  $f_5^k(k=2,3,4)$ can be seen in Fig.1(a). Here,
$VU(5,2,x,r)$ $=$ $(U^\dag)^{(\sum_ix_i) \mod 2}$ $V^{r_5}U_2^{x_5}$ $V^{r_5}U_2^{x_5}$ $V^{r_4}U_2^{x_4}$ $V^{r_3}U_2^{x_3}$ $ V^{r_2}$ $U_2^{x_2}$ $V^{r_1}$ $U_2^{x_1}$ $=$ $U_2^\dag$ $V^0U_2^1$  $V^1U_2^0$  $V^0U_2^0$  $V^0U_2^1$  $V^1U_2^1$ $=$ $U_2^\dag$ $IU_2$  $VI$  $II$  $IU_2$  $VU_2$.
Similarly, $VU(5,3,x,r)$ $=$ $(U_3^\dag)^0$ $IU_3$  $VI$  $II$  $IU_3$  $VU_3$ and
 $VU(5,4,x,r)$ $=$ $(U_4^\dag)^3$ $IU_4$  $VI$  $II$  $IU_4$  $VU_4$.
The measurement results can be seen in Fig.2, and the probabilities of output $f_5^k(k=2,3,4)$ correctly can be seen in Table\ref{Statistics of output $f_n^k$ correctly with $n=5$}.

 \begin{figure}[tbp]\centering
\includegraphics[height=5cm, width=7.5cm]{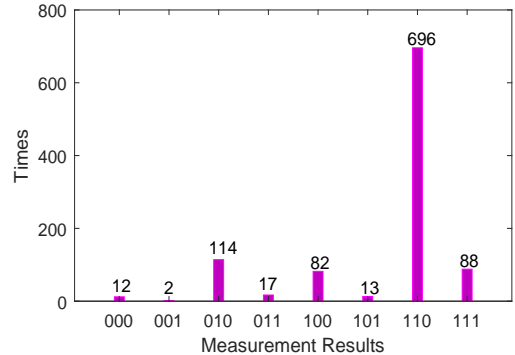}
\caption{(Color online) Measurement results ($f_5^4$, $f_5^3$, $f_5^2$):
 $(0,0,0)$ occured 12 times, $(0,0,1)$ occured 82 times,  $\cdots$ , $(1,1,1)$ occured 88 times and a total of 1024 times.}
\label{Measurement results with n=5}
\end{figure}

\begin{table}
\renewcommand{\arraystretch}{1.9}
\setlength{\abovecaptionskip}{2pt}
\setlength{\belowcaptionskip}{6 pt}
 \centering
 \caption{~~Statistics of output $f_5^k$ correctly} \label{Statistics of output $f_n^k$ correctly with $n=5$}
\begin{tabular}{cccc}
\hline
Boolean function & Correct value  & times & Probability  \\
 \hline
 $f_5^2$ & 1  & 879  & 85.84\% \\
 $f_5^3$ & 1  & 915  & 89.36\%\\
 $f_5^4$ & 0  & 904  & 88.28\%\\
 \hline
\end{tabular}
\end{table}

Case 2: $n=8$. let $C_1, $ $C_2, $ $\cdots, $ and $C_8$ be the eight clients. They will collaborate to compute the function $f_8^2,$ $f_8^3,$ $f_8^4,$ $f_8^5,$ and $f_8^6$ with  $x=(x_1, x_2, \cdots, x_8)$ $=$ $(1, 0, 1,  1, 1, 0, 1, 0)$, $r=(r_1, r_2, \cdots, r_8)$ $=$ $(1, 1, 1,  0, 0, 1, 0, 0)$, $(\sum_i^8x_i) \mod 2=1$, $(\sum_i^8x_i) \mod 3=2$, $(\sum_i^8x_i) \mod 4=1$, $(\sum_i^8x_i) \mod 5=0$, $(\sum_i^8x_i) \mod 6=5$ and $\bar{r}=\oplus_{i=1}^8r_i=0$. The quantum implementation of  $f_8^k(k=2,3,4,6,6)$ can be seen in Fig.1(b). Here,
$VU(8,2,x,r)$ $=$ $U_2^\dag    $  $II$ $IU_2$  $VI$ $IU_2$ $IU_2$  $VU_2$  $VI$  $VU_2$,
$VU(8,3,x,r)$ $=$ $(U_3^\dag)^2$  $II$ $IU_3$  $VI$ $IU_3$ $IU_3$  $VU_3$  $VI$  $VU_3$,
$VU(8,4,x,r)$ $=$ $U_4^\dag    $  $II$ $IU_4$  $VI$ $IU_4$ $IU_4$  $VU_4$  $VI$  $VU_4$,
$VU(8,5,x,r)$ $=$ $(U_5^\dag)^0$  $II$ $IU_5$  $VI$ $IU_5$ $IU_5$  $VU_5$  $VI$  $VU_5$, and
$VU(8,6,x,r)$ $=$ $(U_6^\dag)^5$  $II$ $IU_6$  $VI$ $IU_6$ $IU_6$  $VU_6$  $VI$  $VU_6$.
The measurement results ($f_8^2\oplus \bar{r}$, $f_8^3\oplus \bar{r}$, $f_8^4\oplus \bar{r}$, $f_8^5\oplus \bar{r}$, $f_8^6\oplus \bar{r}$)$=$ ($f_8^2$, $f_8^3$, $f_8^4$, $f_8^5$, $f_8^6$) can be seen in Fig.3, and the probabilities of output $f_8^k(k=2,3,4,5,6)$ correctly can be seen in Table\ref{Statistics of output $f_n^k$ correctly with $n=8$}.

\begin{figure}[tbp]\centering
\includegraphics[height=5cm, width=7.5cm]{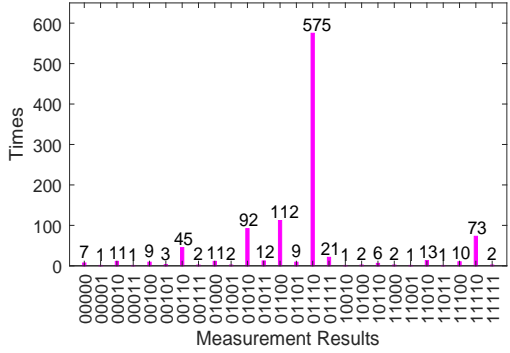}
\caption{(Color online) Measurement results ($f_{8}^6\oplus 1$, $f_{8}^5\oplus 1$,$f_{8}^4\oplus 1$,$f_{8}^3\oplus 1$,$f_{8}^2\oplus 1$,):
 $(0,0,0,0,0)$ occured 7 times, $(0,0,0,0,1)$ occured 1 times,  $\cdots$ , $(1,1,1,1,1)$ occured 2 times and a total of 1024 times. Besides, the results $(1,0,0,0,1)$,$(1,0,0,1,1)$,$(1,0,1,0,1)$, $(1,0,1,1,1)$ and $(1,1,1,0,1)$ did not appear.}
\label{Measurement results with n=8}
\end{figure}

\begin{table}
\renewcommand{\arraystretch}{1.9}
\setlength{\abovecaptionskip}{2pt}
\setlength{\belowcaptionskip}{6 pt}
 \centering
 \caption{~~Statistics of output $f_8^k$ correctly} \label{Statistics of output $f_n^k$ correctly with $n=8$}
\begin{tabular}{cccc}
\hline
Boolean function & Correct value  & times & Probability  \\
 \hline
 $f_8^2$ & 0  & 969  & 94.63\% \\
 $f_8^3$ & 1  & 855  & 83.50\%\\
 $f_8^4$ & 1  & 869  & 84.86\%\\
 $f_8^5$ & 1  & 936  & 91.41\%\\
 $f_8^6$ & 0  & 913  & 89.16\%\\
 \hline
\end{tabular}
\end{table}

Case 3: $n=10$. let $C_1, $ $C_2, $ $\cdots, $ and $C_{10}$ be the eight clients. They will collaborate to compute the function $f_{10}^2,$ $f_{10}^3,$ $f_{10}^4,$ $f_{10}^5,$ and $f_{10}^6$ with  $x=(x_1, x_2, \cdots, x_{10})$ $=$ $(1, 0, 1, 0, 1, 0, 1, 0, 1, 0)$, $r=(r_1, r_2, \cdots, r_{10})$ $=$ $(1, 0, 1, 1,  0, 0, 1, 0, 1, 0)$, $(\sum_i^{10}x_i) \mod 2=1$, $(\sum_i^{10}x_i) \mod 3=2$, $(\sum_i^{10}x_i) \mod 4=1$, $(\sum_i^{10}x_i) \mod 5=0$, $(\sum_i^{10}x_i) \mod 6=5$ and $\bar{r}=\oplus_{i=1}^{10}r_i=1$.
The quantum implementation of  $f_{10}^k(k=2,3,4,5,6)$ can be seen in Fig.1(c). Here,
$VU(10,2,x,r)$ $=$ $U_2^\dag    $  $II$ $VU_2$ $II$ $VU_2$ $II$  $IU_2$ $VI$ $VU_2$ $II$ $VU_2$,
$VU(10,3,x,r)$ $=$ $(U_3^\dag)^2$  $II$ $VU_3$ $II$ $VU_3$ $II$  $IU_3$ $VI$ $VU_3$ $II$ $VU_3$,
$VU(10,4,x,r)$ $=$ $U_4^\dag    $  $II$ $VU_4$ $II$ $VU_4$ $II$  $IU_4$ $VI$ $VU_4$ $II$ $VU_4$,
$VU(10,5,x,r)$ $=$ $(U_5^\dag)^0$  $II$ $VU_5$ $II$ $VU_5$ $II$  $IU_5$ $VI$ $VU_5$ $II$ $VU_5$, and
$VU(10,6,x,r)$ $=$ $(U_6^\dag)^5$  $II$ $VU_6$ $II$ $VU_6$ $II$  $IU_6$ $VI$ $VU_6$ $II$ $VU_6$.
The measurement results ($f_{10}^6\oplus \bar{r}$, $f_{10}^5\oplus \bar{r}$, $f_{10}^4\oplus \bar{r}$, $f_{10}^3\oplus \bar{r}$, $f_{10}^2\oplus \bar{r}$)$=$ ($f_{10}^6\oplus 1$, $f_{10}^5\oplus 1$, $f_{10}^4\oplus 1$, $f_{10}^3\oplus 1$, $f_{10}^2\oplus 1$) can be seen in Fig.4, and the probabilities of output $f_8^k(k=2,3,4,5,6)$ correctly can be seen in Table\ref{Statistics of output $f_n^k$ correctly with $n=8$}.

 \begin{figure}[tbp]\centering
\includegraphics[height=5cm, width=7.5cm]{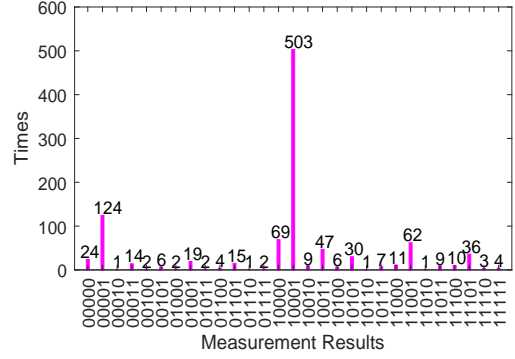}
\caption{(Color online) Measurement results ($f_{10}^6\oplus 1$, $f_{10}^5\oplus 1$,$f_{10}^4\oplus 1$,$f_{10}^3\oplus 1$,$f_{10}^2\oplus 1$,):
 $(0,0,0,0,0)$ occured 24 times, $(0,0,0,0,1)$ occured 124 times,  $\cdots$ , $(1,1,1,1,1)$ occured 4 times and a total of 1024 times. Besides, the results $(0,0,1,0,1)$, $(0,0,1,1,1)$ and $(0,1,0,1,0)$ did not appear.}
\label{Measurement results with n=10}
\end{figure}

\begin{table}
\renewcommand{\arraystretch}{1.9}
\setlength{\abovecaptionskip}{2pt}
\setlength{\belowcaptionskip}{6 pt}
 \centering
 \caption{~~Statistics of output $f_{10}^k$ correctly } \label{Statistics of output $f_n^k$ correctly with $n=10$}
\begin{tabular}{cccc}
\hline
Boolean function & Correct value  & times & Probability  \\
 \hline
 $f_{10}^2$ & 0  & 880  & 85.94\% \\
 $f_{10}^3$ & 1  & 923  & 90.14\%\\
 $f_{10}^4$ & 1  & 897  & 87.60\%\\
 $f_{10}^5$ & 1  & 843  & 82.32\%\\
 $f_{10}^6$ & 0  & 808  & 78.91\%\\

 \hline
\end{tabular}
\end{table}

\section{Conclusion}
In this paper, we solves an open problem  proposed by Clementi et al\cite{CPE2017}. Inspired by Clementi et al's work, we explore the quantum realization of symmetric Boolean functions and  demonstrate that a class of $n-variable$ symmetric Boolean functions $f_n^k$ can be implemented by quantum circuits. Besides, each function $f_n^k(2\leq k\leq n)$ can be used to perform secure multiparty computation and each $n-variable$ symmetric Boolean function can be represented by the linear combination of $f_n^k(k=0,1,\cdots,n)$. Also, we propose an universal quantum implementation method for arbitrary $n-variable$ symmetric Boolean function $f(x_1, x_2, \cdots, x_n)$. At last, we demonstrate our secure multiparty computation protocol on IBM quantum cloud platform.

\acknowledgements
This work was supported in part by the National Key R\&D Program of China under Grant 2017YFB0802400, the National Science Foundation of China under Grant 61373171, 61702007 and 11801564, the 111 Project under Grant B08038, and the Key Project of Science Research of Anhui Province under Grant KJ2017A519, the Natural Science Foundation of Shaanxi Province under Grant No.2017JQ1032, the Basic Research Project of Natural Science of Shaanxi Province under Grant 2017JM6037.


\begin{thebibliography}{99}

\bibitem{Y1982}
A. C.Yao,
 in proceedings of the 23rd Annual Symposium on Foundations of Computer Science(IEEE Computer Society, Washington, DC,1982), pp.160-164.

\bibitem{Y1986}
A. C.Yao,
 in proceedings of the 27rd Annual Symposium on Foundations of Computer Science(IEEE Computer Society, Toronto, 1986), pp.162-167.

\bibitem{OS1987}
O. Goldreich, S. Micali, and A. Wigderson, in Annual ACM Symposium on Theory of Computing (ACM, New York, 1987), pp.218.

\bibitem{BB1984} C. H.Bennett and G.Brassard,
 in proceedings of International Conference on Computer System and Signal Processing (Bangalore, India, 1984), pp.175-179.

 \bibitem{HBB1999}
 M. Hillery, V. Buzek and A. Berthiaume,
    Phys. Rev. A {\bf 59}, 1829(1999).

 \bibitem{CM2017}
  H. Cao and W. Ma,
  IEEE. Photonics. J {\bf 9}, 7600207(2017).

\bibitem{ZDS2017}
 W. Zhang, D. S. Ding, Y. B. Sheng, et al,
 Phys. Rev. Lett {\bf 118}, 220501(2017).

\bibitem{CM2018}
H. Cao and W. Ma,
 Laser. Phys. Lett {\bf 15}, 095201(2018).

 \bibitem{TLH2012}
 H. Y. Tseng, J. Lin and T. Hwang,
 Quantum. Inf. Process {\bf 11}, 373(2012).

 \bibitem{AB2009}
 J. Anders and D. E. Browne,
 Phys. Rev. Lett {\bf 102}, 050502(2009).

 \bibitem{LB2010}
 K. Loukopoulos and D. E. Browne,
  Phys. Rev. A {\bf 81}, 062336(2010).

\bibitem{DKK2016}
V. Dunjko, T. Kapourniotis, and E. Kashefi,
J. Quantum. Inform. Comput {\bf 16}, 0061 (2016).

\bibitem{BDS2016}
S. Barz, V. Dunjko, and F. Schlederer, et al,
Phys. Rev. A {\bf 93}, 032339 (2016).

\bibitem{CPE2017}
 M. Clementi, A. Pappa, and A. Eckstein, et al,
Phys. Rev. A {\bf 96}, 062317(2017).

\bibitem{HEK2011}
 Y. Huang, D. Evans, and J. Katz, et al,
in Proceedings of the 20th USENIX Security Symposium, (USENIX,  San Francisco, CA, 2011), pp. 331-335.




\end{thebibliography}
\end{document}